\documentstyle[12pt,equation]{article}
\begin{document}
\newcommand{\be}{\begin{equation}}
\newcommand{\ben}{\begin{subequations}}
\newcommand{\een}{\end{subequations}}
\newcommand{\beq}{\begin{eqalignno}}
\newcommand{\eeq}{\end{eqalignno}}
\newcommand{\ee}{\end{equation}}
\newcommand{\wt}{\widetilde}
\newcommand{\mchi}{\mbox{$m_{\chi}$}}
\newcommand{\lsp}{\mbox{$\tilde {\chi}_1^0$}}
\newcommand{\Ochi}{\mbox{$\Omega_{\tilde \chi} h^2$}}
\newcommand{\tanb}{\mbox{$\tan \! \beta$}}
\newcommand{\cotb}{\mbox{$\cot \! \beta$}}
\renewcommand{\thefootnote}{\fnsymbol{footnote}}
\pagestyle{empty}
\begin{flushright}
APCTP 98--004 \\
April 1998 \\
\end{flushright}

\vspace*{2.5cm}
\begin{center}
{\Large \bf Particle Dark Matter Physics: An Update}\footnote{Invited talk
at the {\it Fifth Workshop on Particle Physics Phenomenology}, Pune, India,
January 1998.}  \\
\vspace{5mm}
Manuel Drees \\
\vspace{5mm}
{\it APCTP, 207--43 Cheongryangri--dong, Tongdaemun--gu,
Seoul 130--012, Korea} \\
\end{center}
\vspace{10mm}

\begin{abstract}
This write--up gives a rather elementary introduction into particle
physics aspects of the cosmological Dark Matter puzzle. A fairly 
comprehensive list of possible candidates is given; in each case the
production mechanism and possible ways to detect them (if any) are
described. I then describe detection of the in my view most promising
candidates, weakly interacting massive particles or WIMPs, in
slightly more detail. The main emphasis will be on recent developments.

\end{abstract}
\clearpage
\setcounter{page}{1}
\pagestyle{plain}
\section*{1) Introduction: The Need for Exotic Dark Matter}

Dark Matter (DM) is, by definition, stuff that does not emit
detectable amounts of electromagnetic radiation. At present its
existence can therefore only be inferred \cite{1} from the
gravitational pull it exerts on other, visible, celestial bodies. The
best evidence of this kind comes from the study of galactic rotation
curves. Here one measures the velocity with which globular stellar
clusters, gas clouds, or dwarf galaxies orbit around (other) galaxies,
including our own Milky Way. If the mass of these galaxies was
concentrated in their visible parts, the orbital velocity at large
radii $R$ should decrease like $1/\sqrt{R}$. Instead, in nearly all
cases on finds that it remains approximately constant out to the
largest radius where it can be measured. This implies that the total
mass $M(R)$ felt by an object at radius $R$ must increase linearly
with $R$. Studies of this type imply that 90\% or more of the mass of
(large) galaxies is dark; this is a lower bound, since it is not known
where the growth $M(R) \propto R$ cuts off (as it must, since the
total mass of a galaxy is obviously finite).

Cosmologists like to express mass densities averaged over the entire
visible Universe in units of the critical density $\rho_c \sim 
10^{-29}$g/cm$^3$; the dimensionless ratio is then called $\Omega$, with
$\Omega = 1$ corresponding to a flat Universe. Analyses of galactic
rotation curves imply
\be \label{e1}
\Omega \geq 0.1.
\ee
Studies of larger (than galactic) structures tend to favor larger values
of the total mass density of the Universe. These include the ``weighing''
of (super)clusters of galaxies through (weak) gravitational lensing as
well as through measurements of their X--ray temperature\footnote{These
two methods give somewhat different results. Gravitational lensing seems
to be a more direct, and hence more reliable, measure of the total mass
of a (super)cluster; it favors larger values of $\Omega$.}, as well as
studies of the large--scale streaming of galaxies. A widely, although
not quite universally, accepted lower bound on the total mass density is
\be \label{e2}
\Omega \geq 0.3.
\ee
Note that a putative cosmological constant would not be detectable through
such studies, since it does not affect local gravitational fields. Finally,
naturalness arguments and (most) inflationary models prefer $\Omega = 1.0$
to high accuracy.

Very recently there have been some claims \cite{2} that observations
of type--I supernovae at high redshift $z$ disfavor matter density
$\Omega = 1$ ``at the 95\% confidence level''. Much of this evidence
seems to rest on the observation of three supernovae with $z$ between
0.5 and 1.0.  Type--I supernovae are assumed to be all more or less
the same (``standard bombs''), in which case their absolute distance
can be computed fairly easily from their apparent brightness. Spectral
lines allow to determine the redshift precisely. Observations of these
three supernovae using the Hubble Space Telescope give $\Omega = -0.1
\pm 0.5$ for vanishing cosmological constant, so $\Omega = 1.0$ seems
to be more than two standard deviations away from the central
value. However, if we believe the bound (\ref{e2}), a good part of the
region favored by this recent measurement is excluded. Besides, in
cosmology the errors are nearly always dominated by systematics, and
can therefore not be interpreted in a straightforward statistical
manner.\footnote{There is also a long--running program of
ground--based searches for high$-z$ type--I supernovae. Recent results
also seem to favor $\Omega < 1$ in the absence of a cosmological
constant, but I have not been able to find more details in several
hours of Web browsing.} In my view it is therefore premature to give
up on a total mass density $\Omega = 1.0$, and even more premature to
take Einstein's self--described ``worst mistake'', a (tiny but
non--zero) cosmological constant, seriously. Finally, the requirement
that the Universe be at least 10 billion years old (a conservative lower
limit on the age of the oldest stars) implies
\be \label{upper}
\Omega h^2 \leq 1,
\ee
where $h$ is the (present) Hubble parameter in units of 100 
km/(sec$\cdot$Mpc). Recent observations indicate $h = 0.65 \pm 0.15$.

The total density of luminous matter only amounts to less than 1\% of
the critical density \cite{1}. Moreover, analyses of Big Bang
nucleosynthesis determine the total baryonic density to lie in the
range \cite{2a}
\be \label{e3}
0.01 \leq \Omega_{\rm baryon} h^2 \leq 0.015.
\ee
Assuming $h \geq 0.5$, the upper bound in (\ref{e3}) implies 
\be \label{e3a}
\Omega_{\rm baryon} \leq 0.06,
\ee
in mild conflict with the lower bound (\ref{e1}),\footnote{There has
been a recent claim in the literature \cite{3a} that a modification
of the assumed mass--to--light ratio would allow to explain galactic
rotation curves without introducing exotic DM; this amounts to the
assumption that the outer parts of galaxies contain many more small,
and hence dim, stars than one finds in the solar neighbourhood. Another
recent article \cite{3b} claims to have found indirect evidence for the
existence of fairly compact, partly ionized gas clouds in the halo of
our galaxy, based on the apparent modification of some radio waves emitted
by quasars. Finally, yet a third paper \cite{3c} claims to have found
some indication that the dark halo of our galaxy consists of Jupiter--sized
primordial black holes. Note that, unlike the first two suggestions, this
would allow to circumvent the bound (\ref{e3a}) if these black holes
were formed before the era of nucleosynthesis. It is quite possible, even
likely, that these claims will be refuted in due course. Nevertheless one
should keep in mind that purely astrophysical solutions may yet contribute
to the solution of (part of) the DM puzzle.} and in blatant
conflict with (\ref{e2}). Most Dark Matter must therefore be 
non--baryonic.\footnote{Note that the lower bound in (\ref{e3}) implies
$\Omega_{\rm baryon} \geq 0.015$; hence BBN not only allows for, but
actually demands, the existence of baryonic DM. Currently MACHOs
observed in microlensing experiments \cite{3} seem to be the best
candidates for this.} This, of course, is where particle physics comes
into play; some sort of ``new physics'' seems to be required to 
describe this exotic matter, beyond the particles described by the Standard
Model of particle physics (SM). According to the best (post--MACHO)
estimate \cite{4}, the local density of this mysterious stuff amounts to
about
\be \label{e4}
\rho_{\rm local}^{\rm DM} \simeq 0.3 \ {\rm GeV/cm}^3
\simeq 5 \cdot 10^{-25} {\rm g/cm}^3.
\ee
It is assumed to have a Maxwellian velocity distribution with mean
$\bar v \simeq 300$~km/sec. The local flux of DM ``particles'' $\chi$
(which could be very massive, e.g. black holes) is thus
\be \label{e5}
\Phi_{\rm local}^{\rm DM} \simeq \frac {100 \ {\rm GeV}} {\mchi}
\cdot 10^5 \ {\rm cm}^{-2} {\rm s}^{-1}.
\ee

The task for particle physicists is two--fold. First, one has to
identify a particle (or configuration of fields) that has the proper
universal relic density. Second, one should find a way to detect these
putative relics from the early Universe. As we will see in the following
two sections, the first task is easily accomplished; in fact, there is
an ``embarrassment of riches'' of particle DM candidates. Testing these
hypotheses experimentally, however, may well prove very difficult; in some
cases it seems basically impossible.

\section*{2) Dark Matter Candidates}

Particle Physicists have come up with a fairly long list of DM candidates.
Many of them have been described in other recent reviews \cite{5}. I
therefore treat these ``usual suspects'' comparatively briefly, and
instead put somewhat more emphasis on developments in the last year or
so.

\setcounter{footnote}{0}
\subsection*{2a) Light Neutrinos}
A light ($m_{\nu} \leq$ MeV) neutrino contributes \cite{5a}
\be \label{e6}
\Omega_{\nu} h^2 \simeq m_\nu / (90 \ {\rm eV})
\ee
to the scaled mass density of the Universe; that is, for $h \simeq 0.6$,
neutrinos give $\Omega \simeq 1$ if $\sum_i m_{\nu_i} \simeq 30$~eV. This
does not violate any direct experimental limits if most of the mass 
comes from $\mu$ or $\tau$ neutrinos. However, evidence for neutrino
oscillations suggests that mass differences between the three known
neutrinos are of order 1~eV or less \cite{6}. Together with the upper bound
on the $\nu_e$ mass from studies of the tritium $\beta-$decay spectrum 
this seems to indicate that the three known neutrinos cannot contribute 
$\Omega_{\nu} \simeq 1$. 

Moreover, light neutrinos constitute ``hot'' DM, i.e. they were still
relativistic at the onset of galaxy formation (after the recombination of
electrons and ions, i.e. after the freeze--out of the cosmic background
radiation, about $3 \cdot 10^5$ years after the Big Bang). This seems
incompatible with the observed pattern of structure in the Universe,
{\em if} density inhomogeneities were seeded by quantum fluctuations during
the (hypothetical) inflationary epoch. However, neutrino DM still seems
to be able to describe structure formation if it is seeded by cosmic
strings \cite{7}; and even in the standard inflationary scenario a
hot$+$cold mixture might describe structure formation better than
pure cold DM, if the hot component amounts to 20--30\% of all DM \cite{1}.
It may therefore be premature to exclude neutrinos as (a component of) DM.

Unfortunately light relic neutrinos seem to be almost impossible to detect.
Studies of $\mu$ and $\tau$ decays are not likely to ever tell us whether
the $\mu$ or $\tau$ neutrino carry a mass of a few (tens of) eV. If
neutrinos mix, oscillation experiments should eventually be able to tell
us the differences of squared neutrino masses; as mentioned above, current
evidence suggests that these differences are small. In this case careful
studies of tritium $\beta$ decay and of neutrinoless double beta decay
(e.g. of ${}^{76}$Ge) should soon tell us if neutrinos can contribute
significantly to DM, {\em if} there are only three neutrinos. If sterile
neutrinos exist, essentially all bets are off. In this case the only hope
seems to be to wait for the next nearby (but not too close, one hopes)
supernova; precise measurements of the time structure of the neutrino
pulse ought to be able to determine the masses of active neutrinos to a few
eV at least \cite{8}.

\subsection*{2b) Axions}

Axions \cite{9} are neutral spin--0 CP--odd (pseudoscalar) particles
associated with the spontaneous breakdown of a new global $U(1)_{PQ}$
symmetry introduced by Peccei and Quinn as a solution of the strong CP
problem. The basic idea is that the CP violating phase $\theta_{\rm
QCD}$ is transformed into into a dynamical variable; the axion
potential is then chosen such that it is minimized for $\theta_{\rm
QCD} =0$. The axion would be massless except for chiral symmetry
breaking in QCD, which gives it a mass of order
\be \label{e7}
m_a \simeq 0.6 \ {\rm meV} \cdot 10^{10} \ {\rm GeV} / f_a.
\ee
Here, $f_a$ is the scale where the PQ symmetry is broken. Note that
not only the mass of the axion but also its couplings to ordinary
matter are proportional to $1/f_a$. A combination of laboratory and
cosmological constraints therefore leads to the lower bound \cite{9}
\be \label{e8}
f_a \geq 5 \cdot 10^9 \ {\rm GeV}.
\ee
On the other hand, the axions' contribution to the total mass of the
Universe increases with $f_a$. The requirement $\Omega h^2 \leq 1$
therefore implies \cite{11}
\be \label{e9}
f_a \leq 10^{12} \ {\rm GeV}.
\ee
This upper bound is somewhat soft. If the Universe (after inflation)
ever had a temperature exceeding $f_a$, axionic strings should have
formed during the PQ phase transition; these are potent sources of
axions, so the bound (\ref{e9}) would be strengthened. On the other hand,
if the axion field at the onset of the QCD phase transition ``happened''
to have had a small value, this bound could be relaxed considerably.
Certain supersymmetric axion models also allow to relax this bound by
up to three orders of magnitude, if the decay of the superpartners of the
axion produces a sufficient amount of entropy after the epoch of
axion production \cite{12}.

The bound (\ref{e8}) implies that axions interact far too weakly for us to
be able to detect axions produced in laboratory experiments. Indeed, relic
axions seem to be the only ones that might be detectable with current
technology. Note that axions are produced athermally; they behave like
cold DM, and should thus cluster in our galaxy. Also, the bounds
(\ref{e8}) and (\ref{e9}) imply that, if axions exist at all, their
relic density ought to be substantial.

Currently two axion search experiments are under way, one in Japan and
one in the US. Both attempt to convert relic axions into microwave
photons in the presence of a strong magnetic field. The US experiment
plans to cover the range $2 \mu$eV $\leq m_a \leq$ 12~$\mu$eV, with
moderate sensitivity, while the Japanese experiment intends to search
for axions with $m_a \simeq 10 \ \mu$eV with very high sensitivity.
The US experiment has just published their first results \cite{13a},
which exclude one kind of axions with mass between 2.9 and 3.3 $\mu$eV.
As far as I know, the Japanese experiment \cite{13b} has not released any 
results yet.

\subsection*{2c) WIMPs}
Weakly Interacting Massive Particles (WIMPs) are the most widely studied
particle physics DM candidates. Their name derives from a peculiar
``coincidence''. The relic density of massive particles that were already
more or less non--relativistic when they decoupled from the hot bath
of SM particles is approximately (up to logarithmic corrections)
given by \cite{14}
\be \label{e10}
\Omega_{\chi} h^2 \simeq \frac {C} {\langle \sigma_{\rm ann}^{\chi}
\cdot v \rangle },
\ee
where $\sigma_{\rm ann}^{\chi}$ is the total annihilation cross section
of WIMP pairs into SM particles, $v$ is the relative velocity between
the two WIMPs in their cms, and $\langle \dots \rangle$ denotes 
thermal averaging. The constant $C$ involves factors of Newton's
constant, the temperature of the cosmic background radiation, etc. The
``coincidence'' mentioned above is that its numerical value is such that
$\Omega_{\chi} h^2$ comes out in the desired range (of order 0.2) if
the annihilation cross section is roughly of order of weak scattering
cross sections. This hints at a connection between the total mass of the
Universe and (generalized) electroweak physics.

The simplest example for a WIMP would have been a massive ($m_\nu \geq$
GeV) neutrino, but this possibility is now nearly excluded by a 
combination of searches at the $e^+ e^-$ collider LEP (specifically,
by the upper bound on any non--SM contribution to the invisible width of
the $Z$ boson), cosmological arguments, and DM searches (to be described
in the next section); the only loophole would be a massive neutrino that
is mostly (but not entirely) an $SU(2) \times U(1)_Y$ singlet. It would be
very difficult to understand why such a particle should be stable,
however.

The by far best motivated WIMP is the lightest supersymmetric particle,
specifically the lightest neutralino \lsp\ \cite{14}. It is stable by
virtue of $R$ parity, which is automatically conserved if one only
introduces absolutely necessary interactions in the SUSY 
Lagrangian.\footnote{This minimality argument can be made more rigorous
in models with gauged $B-L$ symmetry \cite{15}.} Moreover, there are
several regions of parameter space where $\Omega_{\chi}$ comes
out close to the desired value.

This last point deserves a bit more discussion. Assuming minimal particle
content, as in the MSSM, \lsp\ is in general a linear superposition
of the fermionic superpartners of the $U(1)_Y$ and neutral $SU(2)$
gauge bosons and of the two neutral Higgs bosons; in the notation of
ref.\cite{16}:
\be \label{e11}
\lsp = N_{11} \widetilde{B} + N_{12} \widetilde{W}_3 + N_{13}
\tilde{h}_1^0 + N_{14} \tilde{h}_2^0.
\ee
Not surprisingly, the properties of the LSP depend quite strongly on
which of these four components are large.

A bino-- ($|N_{11}| \simeq 1$) or photino--like ($N_{11} \simeq 
\cos \! \theta_W, \ N_{12} \simeq \sin \! \theta_W$) LSP annihilates
predominantly through the exchange of a sfermion in the $t-$ or
$u-$channel (unless an $s-$channel diagram is ``accidentally'' enhanced;
see below). Due to their large (hyper)charges and, in many models,
comparatively small masses, the exchange of $SU(2)$ singlet sleptons
is especially important. The relic density of a ``generic'' bino--like
LSP can therefore be approximated by \cite{17}
\be \label{e12}
\Omega_{\chi} h^2 \simeq \frac {\Sigma^2} {(1 \ {\rm TeV})^2
m^2_{\chi} } \cdot \frac {1} { \left( 1 - m^2_{\chi}/\Sigma
\right)^2 + m^4_{\chi} / \Sigma^2 },
\ee
where $\Sigma = m^2_{\chi} + m^2_{\tilde e_R}$, and three degenerate
sleptons have been assumed. This yields interesting relic densities for
very reasonable choices of parameters, $\mchi < m_{\tilde e_R} \leq 
350$~GeV and $m_{\tilde e_R} > 100$~GeV. Unfortunately this cannot be
translated into a stringent upper bound on slepton masses, since 
$s-$channel diagrams can greatly increase the annihilation cross section, 
i.e. decrease the predicted relic density. This occurs if the LSP mass
is very close to, or slightly below, half of the mass of one of the
neutral Higgs bosons of the MSSM; such a scenario can be realized even
in the framework of ``minimal Supergravity'' models \cite{17,24}, where one
postulates a very simple form of the sparticle spectrum at an energy
scale not far from the scale of Grand Unification, $M_X \simeq 2 \cdot
10^{16}$~GeV. In such cases LSP masses of several TeV, while extremely
ugly from the point of view of particle physics, cannot be excluded from
cosmology.

The situation is a bit more complicated for a higgsino--like LSP
($|N_{13}|^2 + |N_{14}|^2 \simeq 1$). In the framework of the MSSM this
usually implies $|N_{13}| \simeq |N_{14}| \ (\simeq 1/\sqrt{2})$, 
which means that the $Z\lsp\lsp$ coupling is suppressed (it vanishes
in the limit of exact equality \cite{18}). Since the $\lsp f \tilde{f}$
couplings are now Yukawa couplings, sfermion exchange contributions are
small for LSPs that are too light to annihilate into $t \bar{t}$
pairs. The annihilation cross section for light higgsino--like LSPs
is therefore small.

Their relic density nevertheless turns out to be quite small \cite{19},
due to ``co--annihilation'' \cite{20} between LSPs and heavier
neutralinos or charginos. If \lsp\ is higgsino--like, so are (usually)
$\tilde{\chi}_2^0$ and $\tilde{\chi}_1^\pm$; moreover, the masses of
these three particles are quite close to each other. Recall that \lsp\
is already non--relativistic at freeze--out, i.e. its number density
is suppressed exponentially (by a Boltzmann factor) compared to those
of SM particles. Scattering reactions of the type $\lsp f \leftrightarrow
\tilde{\chi}_2^0 f$ and $\lsp f \leftrightarrow \tilde{\chi}_1^\pm f'$
therefore occured much more frequently
than annihilation reactions like $\lsp \lsp \leftrightarrow f \bar{f}$;
here $f, f'$ denote SM (matter) fermions. As a result, the \lsp, 
$\tilde{\chi}_2^0$ and $\tilde{\chi}_1^\pm$ densities remain in
{\em relative} equilibrium long after superparticles have decoupled from
ordinary particles. This is important, since the $Z \lsp \tilde{\chi}_2^0$
and $W^\pm \lsp \tilde{\chi}_1^\mp$ couplings are unsuppressed if
\lsp\ is higgsino--like. Hence the cross sections for $\lsp 
\tilde{\chi}_2^0$ and $\lsp \tilde{\chi}_1^\pm$ co--annihilation into
SM fermion antifermion pairs are orders of magnitude larger than those
for \lsp\lsp\ annihilation. This greatly reduces the final \lsp\
density, to the extent that only a small sliver of parameter space
survives where a higgsino--like LSP with mass below $M_W$ makes a
good DM candidate; among other things, one has to assume that certain
quantum corrections to the masses of the higgsino--like states
occur with the correct sign \cite{21}.

A higgsino--like LSP with mass $\geq M_W$ has very large annihilation
cross sections into $W$ and $Z$ pairs. Even in the absence of co--annihilation
one therefore needs LSP masses exceeding 400 GeV to get a cosmologically
interesting relic density. (Note that the annihilation cross section
must drop at least as $1/m^2_{\tilde \chi}$, due to partial wave
unitarity.) Moreover, a recent complete calculation of co--annihilation
of heavy higgsino--like states \cite{22} showed that it reduces the 
predicted relic density significantly. Even though annihilation into
$W$ and $Z$ pairs are the biggest single cross sections, the {\em total}
co--annihilation cross sections remain significantly larger than 
annihilation cross sections, due to the large number of different
fermion antifermion final states, which essentially only contribute to the
former. This pushes the lower bound on the mass of a cosmologically
interesting higgsino--like LSP into the TeV region. Such a heavy LSP is
not natural from the particle physics point of view; one would need a
significant amount of finetuning \cite{23} to keep the $W$ and $Z$ bosons
as light as they are, if even the lightest superparticle weighs at least
a TeV. This would weaken the very motivation for ``weak scale''
supersymmetry.\footnote{The authors of ref.\cite{22} also find
examples where co--annihilation {\em increases} the prediction for the LSP
relic density. This can happen if there are three nearly degenerate
neutralino states, i.e. if the supersymmetry breaking bino mass parameter
$M_1$ is close to the supersymmetric higgsino mass parameter $|\mu|$.
Since the cross section for co--annihilation of a gaugino--like and a
higgsino--like state is quite small, conversion of the LSP into a
heavier neutralino can then enhance its probability to survive until
freeze--out. However, this phenomenon only occurs for a narrow strip of
parameter space.}

Finally, if the LSP is $\widetilde{W}_3-$like ($|N_{12}| \simeq 1$), its
predicted relic density is again very small unless the LSP mass
exceeds 1 TeV \cite{25}. The culprit is again co--annihilation; in this
case the mass splitting to the lightest chargino often only amounts to
a few hundred MeV. Fortunately such a situation appears somewhat unlikely
from the model building point of view, since it requires strong violation
of the unification of gaugino mass parameters (at the GUT scale, the
$U(1)_Y$ gaugino mass must be more than twice as large as the $SU(2)$
gaugino mass); such scenarios do exist, however \cite{25a}.

While the lightest MSSM neutralino is the by far most widely studied WIMP
candidate, other examples also have been suggested. In the NMSSM, where
one introduces one $SU(2) \times U(1)_Y$ singlet Higgs superfield, 
the LSP usually is still the lightest neutralino, but it now also has
a ``singlino'' component [i.e., a fifth term must be added to 
eq.(\ref{e11})]. This allows for new solutions with interesting relic
density \cite{26}. Similar remarks apply for models with an extra
$U(1)$ gauge factor \cite{26a}.

Another possible WIMP candidate \cite{27} is the lightest ``messenger
sneutrino'' in models where supersymmetry breaking is mediated to the
visible sector through gauge interactions. Unlike an ``ordinary''
sneutrino, such a particle could evade bounds from direct WIMP searches
(see Sec.~3a) if quantum corrections split the CP--even and CP--odd
components of the sneutrino field (associated with its real and
imaginary part, respectively), so that only one of these components
survives to the present time \cite{28}. However, the messenger scale
would then have to be quite low in order to avoid getting too high a
relic density, in violation of the bound (\ref{upper}).\footnote{One
can still construct models with very high messenger scales if one
arranges for the lightest messenger particle to be unstable, or for it
to be ``inflated away''. However, it can then no longer be a DM candidate.}
Similarly, the CP--even and CP--odd components of $\tilde{\nu}_\tau$
could be split by some new lepton flavor violating interaction
\cite{29}; this mechanism would require a $\nu_\tau$ mass of at least
5 MeV, which might be measurable at the upcoming $B-$factories.

Several schemes for detecting relic WIMPs have been suggested; these
will be discussed in Sec.~3.

\subsection*{2d) Strongly Interacting Particles}

Certain extensions of the SM also contain strongly interacting DM
candidates.  Note that the ``strong'' interactions referred to here
are a new gauge force, not standard QCD. The oldest example of this
kind \cite{30} is a techni--baryon associated with a hypothetical
techni--color (TC) gauge group, whose techni--quark condensates are
supposed to be responsible for electroweak symmetry breaking. In
analogy with QCD one expects the lightest techni--baryon to be stable,
but it can annihilate through TC interactions into unstable
techni--pions. Techni--baryons have masses of roughly a few TeV; given
their strong annihilation cross sections, the standard formalism
described in the previous subsection would then predict too low a
relic density to be of cosmological interest. However, it is possible
that there is a techni--baryon number asymmetry in the Universe. In
fact, mechanisms have been suggested that would generate approximately
equal baryon and techni--baryon asymmetries. In this case one would
expect
\be \label{e13}
\Omega_{TC} \sim \Omega_{\rm baryon} \cdot \frac {m_{TC}} {m_p},
\ee
where $m_p$ is the proton mass. Given that $\Omega_{\rm baryon} > 0.01$,
see eq.(\ref{e3}), this is actually too large by at least one order
of magnitude, so one has to postulate a ``model--specific suppression
factor'' \cite{30}. Note also that 20 years of effort have not produced
a single fully viable techni--color model. Techni--baryons would
resemble heavy WIMPs as far as DM searches are concerned \cite{30}.

Another strongly interacting DM candidate \cite{27} originates in the
``secluded sector'' of SUSY models where supersymmetry is broken at
relatively low energies, ${\cal O}(100 \ {\rm TeV})$. Strongly interacting
``baryons'' of this mass would have $\Omega \sim 1$ if no ``baryon''
asymmetry exists in this sector. Indeed, from partial wave unitarity
one can derive \cite{32} the upper bound
\be \label{e14}
m_X \leq 500 \ {\rm TeV}
\ee
for any particle $X$ whose lifetime exceeds the age of the Universe,
{\em if} this particle has ever been in chemical equilibrium with SM
particles (after inflation). An absolutely stable particle of this
mass is probably all but undetectable \cite{27} if it is a singlet
under the SM gauge group, as is expected for ``secluded baryons''.

Strongly interacting particles always behave like cold DM.

\subsection*{2e) Gravitationally Interacting Particles}

On the other end of the spectrum are DM candidates that only have
gravitational interactions. The most prominent example is the
gravitino \cite{33}. If it is much lighter than ``visible'' sparticles (as
it has to be in order to make a viable DM candidate), its interactions
are actually of ``enhanced gravitational strength'', since the effective
coupling of the spin--1/2 (Goldstino) component of the gravitino is
$s/(m_{3/2} M_{Pl})$, where $s$ is the squared cms energy of the
process under consideration, $m_{3/2}$ is the gravitino mass, and $M_{Pl}$
is the Planck mass. This is sufficient for a keV gravitino to have been
in thermal equilibrium at temperatures $T \geq 100$~GeV. Gravitinos
would have been ultra--relativistic at freeze--out, so the calculation
of their relic density resembles that for light neutrinos \cite{5a}.
One finds \cite{33}
\be \label{e15}
\Omega_{\tilde G} h^2 \simeq \frac {m_{3/2}} {0.85 \ {\rm keV}}.
\ee
These gravitinos would have been ``warm'' at the onset of structure
formation; for most, though not all, purposes this would resemble cold
DM. One can also produce an effectively hot (really, athermal) gravitino
component if a large number of LSPs decay into gravitinos \cite{33}; 
however, one needs sfermion masses well in excess of 1 TeV for this
component to be significant.

To my knowledge nobody has yet found a way to detect relic gravitinos.

\setcounter{footnote}{0}
\subsection*{2f) Superheavy Particles}

Recently there has been renewed interest in DM candidates whose masses
greatly exceed the unitarity bound (\ref{e14}). Recall that this bound
assumed that the new particle once was in chemical
equilibrium\footnote{This means that reactions that change the number
of these heavy particles must have been in equilibrium. This is a far
more restrictive constraint than the requirement that the $X$
particles have the same temperature as the bath of SM particles.} with
SM particles.  Stable particles with mass beyond 500 TeV must
therefore have sufficiently small annihilation cross sections to never
have achieved equilibrium \cite{34}: 
\be \label{e16} 
n_X \langle \sigma_{\rm ann}^X v \rangle < H, 
\ee 
where $H$ is the Hubble parameter and $n_X$ the number density of $X$
particles. Note that this inequality must be satisfied at all times
(after the end of the inflationary epoch, if there was one). In the
framework of ``standard'' inflationary models, and assuming
$\Omega_X = 1$, this implies \cite{34}
\be \label{e17} 
m_X \geq 10^{13} \ {\rm GeV} \cdot \left( \frac {100 \
{\rm GeV}} {T_{RH}} \right)^{1/3} \cdot \left( \frac { \langle
\sigma_{\rm ann}^X v \rangle } {m^2_X} \right)^{2/3}.  
\ee 
Here $T_{RH}$ is the reheating temperature after inflation, which
might have been as high as $10^9$~GeV. Also, the annihilation cross
section could be (much) smaller than the unitarity limit $\sim
1/m^2_X$.  Eq.(\ref{e17}) therefore suggests that particles with
masses exceeding $10^{13}$~GeV certainly never were in thermal
equilibrium, but (much) lighter particles might also satisfy this
criterion.

By definition the $X-$particle could not have formed thermally, but
several other mechanisms have been suggested \cite{34}. If $m_X$ is
less than the mass of the inflaton $m_\phi \simeq 10^{13}$~GeV, it
could be formed in inflaton decays. Even if $m_X > m_\phi$ a 
substantial $X$ density might have been created during a period of
``pre--heating'' through a ``parametric resonance'' \cite{35}. 
Neither of these two mechanisms seems to particularly favor
$\Omega_X \sim 1$, though. In contrast, (gravitational) interactions
between $X-$particles and the space--time metric at the end of inflation
naturally produce \cite{34} $\Omega_X \sim 1$ (to logarithmic accuracy,
anyway) if $m_X \sim m_\phi \ (\sim 10^{13} \ {\rm GeV})$. Finally, $X$
might be the inflaton itself \cite{35a}.

Of course, $X$ also has to be very long--lived for it to be a good DM
candidate. In particular, its couplings to SM particles must be far
weaker than gravity \cite{36}. This probably excludes massive
Kaluza--Klein modes, which are expected to exist in superstring (or $M$)
theory. On the other hand, $10^{13}$~GeV is close to the scale where
SUSY is supposed to be broken in the hidden sector of many supergravity
(or superstring) theories. Some ``baryons'' in this hidden sector may be
sufficiently long--lived; an explicit example has been suggested in
ref.\cite{36}.

Stable $X-$particles of this mass are almost certainly undetectable. On
the other hand, if their lifetime is roughly comparable to the age of
the Universe, their decay products might be detectable. In fact, it has
been suggested \cite{37} that these have been seen already, in form of
ultra--high energy cosmic rays ($E \geq 10^{11}$~GeV)!\footnote{The origin of
these extremely energetic events is very difficult to explain in the 
framework of the SM.}

\setcounter{footnote}{0}
\subsection*{2g) Supersymmetric $Q-$Balls}

$Q-$balls \cite{38} are stable scalar field configurations carrying a
global conserved $U(1)$ charge $Q$. Such objects might exist in
supersymmetric models \cite{39}, provided the potential is
sufficiently flat for large field values; the $U(1)$ charge in question
would be the baryon number $B$. The MSSM has many directions in field
space that are both $D-$ and $F-$flat (i.e. where the potential feels
neither gauge nor Yukawa interactions), but SUSY breaking has to be
switched off at sufficiently low energies to make the soft breaking
terms sufficiently small for large values of the fields. This
scenario can therefore only work in models with gauge mediated SUSY
breaking, where the soft breaking terms rapidly disappear at scales
above the mass of the messenger fields.\footnote{In gravity mediated
scenarios, $B-$balls might still be sufficiently long--lived to play a
role in the creation of the baryon asymmetry of the Universe
\cite{40}.}  In these models the total mass of the $B-$balls only
grows $\propto B^{3/4}$; hence a $B-$ball becomes stable against decay
into nucleons if 
\be \label{e18} 
B > \left( \frac { m_{\tilde q} } {m_p } \right)^4 
\simeq 10^{10} \cdot \left( \frac { m_{\tilde q} } 
{300 \ {\rm GeV} } \right)^4.  
\ee 
This implies that their mass would have to be at least $10^9$ GeV or
so.  Kusenko and Shaposhnikov argue \cite{39} that such objects could
have been formed abundantly in the decay of a squark field condensate,
which is an ingredient of the Affleck--Dine mechanism \cite{41} of
baryogenesis. At present it is not clear why this should produce
$\Omega \sim 1$, though.

Although fairly heavy, $B-$balls are actually easy to detect
\cite{42}, provided only that their flux (\ref{e5}) is high enough for
one of them to pass through your detector. If the field configuration is
such that electrons have a large mass inside the $B-$ball, it will collect
a large electric charge by ``eating'' protons through the reaction
$q q \rightarrow \tilde{q} \tilde{q}$. Eventually the Coulomb barrier will
become too high and this reaction will stop. A fully charged $B-$ball will
interact very strongly with matter through atomic collisions, releasing
$\sim 100$~GeV of energy on each cm of its track. This has to come from
the kinetic energy of the $B-$ball, i.e. eventually it will get stuck in
matter. Charged $B-$balls with $B \leq 10^{13}$ would therefore not
reach detectors placed deeper than 1 km water--equivalent \cite{42}. 
Heavier ones should reach those detectors; their flux is thus limited
by unsuccessful searches for highly charged ``quark nuggets''. On the
other hand, if electrons are light or massless inside a $B-$ball, it
can shed its electric charge through $u e \rightarrow d \nu_e$. Such a 
neutral $B-$ball also releases $\sim 100$~GeV/cm$\cdot \rho/(1 {\rm 
g/cm}^3)$, 
but this energy comes from ``eating'' nucleons, and is released mostly in
form of pions. Such an object would therefore look like a magnetic
monopole catalyzing nucleon decay as it moves through a detector, so the
corresponding search limits apply. Altogether one has \cite{42}
\ben \label{e19} \beq
B &\leq 10^{13} \ \ {\rm or} \ \ B \geq 10^{21}, \ \ {\rm for \
electrically \ charged} \ B \ {\rm balls}; \label{e19a} \\
B &\geq 3 \cdot 10^{22}, \hspace*{2.4cm} {\rm for \ electrically
\ neutral} \ B \ {\rm balls}.
\eeq \een

Finally, it has been proposed that $B-$balls might be eating up neutron
stars \cite{43}. Eventually an attacked $n-$star would become too light
to remain gravitationally bound, and would be blown apart by its Fermi
pressure. This might give rise to the observed gamma ray bursters.

\section*{3) WIMP Detection}

WIMPs are the in my view best motivated, and certainly most widely
studied, particle DM candidates. I will therefore describe proposals for
their detection in somewhat more detail than for the other candidates
mentioned in the previous Section. Basically three kinds of WIMP signatures
have been discussed in the literature \cite{14}: Direct detection through
elastic WIMP--nucleus scattering; WIMP annihilation in the center of
the Earth or Sun producing high--energy neutrinos; and WIMP annihilation
in the halo of our galaxy producing hard gamma rays, positrons, or
antiprotons. These methods will be described in the following three
subsections.

\subsection*{3a) Direct WIMP Detection}

Here one searches simply for the elastic scattering of a WIMP on a 
nucleus inside a detector; more exactly, one looks for the recoil of
the struck nucleus. Recall that WIMPs in the halo are supposed to have
velocities of about $10^{-3} c$.\footnote{Cowsik et al. have argued
\cite{44} that the true velocity might be at least two times higher. This
has been criticized in refs.\cite{45}. The issue does not yet seem to be
entirely settled, though \cite{46}.} The recoil energy of a nucleus with
mass $m_A$ can therefore at most be 
\be \label{recoil}
E_{\rm recoil}^{\rm max} = 2 v^2_{\chi} m_A \frac { m^2_{\chi}}
{ \left( m_A + m_\chi \right)^2 }.
\ee
Hence typical recoil energies are $10^{-6} m_A$ or less, in the (tens of)
keV range. Detecting this low an energy is far from easy. One has to
use ultrapure materials for both the detector and its immediate
shielding to minimize backgrounds from $\alpha$ and $\beta$ decays.
Most advanced detection schemes in addition (plan to) use methods to
distinguish between nuclear recoils and $\beta$ and $\gamma$ backgrounds,
e.g. by using pulse shape information (in scintillator detectors \cite{47}),
or by measuring both phonon and ionization energy (in cold
semiconductor detectors \cite{48}).

The first experiments of this type \cite{49} have already achieved sufficient
sensitivity to exclude massive Dirac neutrinos, or (complex) sneutrinos,
as a major component of the galactic halo. The sensitivity limit of these
early experiments was tens or hundreds of events per kg and day. The
currently best bound has been established by the DAMA collaboration
\cite{50}, which operates 9 NaI crystals in the Gran Sasso underground
laboratory; it improves the bound of refs.\cite{49} by up to a factor of 5.

Unfortunately the predicted counting rate for the most attractive WIMP
candidate, the lightest MSSM neutralino, is usually quite low. LSPs can
interact with nucleons through the exchange of squarks in the $s-$ or
$u-$channel, or through $t-$channel exchange of a $Z$ or neutral Higgs
boson \cite{14}. Squark exchange is suppressed by the large squark
mass ($m_{\tilde q} \geq 200$~GeV from searches at the Tevatron $p \bar p$
collider \cite{51}). The exchange of $Z$ bosons is suppressed by the smallness
of the $Z \lsp \lsp$ coupling. Moreover, $Z$ exchange only produces a
spin--spin coupling, since the vector current of the Majorana 
neutralino vanishes identically. Unlike scalar (spin--independent) couplings,
such spin--dependent couplings are not enhanced for heavy nuclei, compared
to the coupling to a single nucleon.

This leaves Higgs exchange as the (usually) most important contribution
to the LSP--nucleus scattering matrix element. The squared Higgs
exchange contribution scales like the inverse fourth power of the mass
of the corresponding Higgs boson;\footnote{Note that the contribution from
exchange of the CP--odd Higgs boson is negligible, since the
pseudoscalar LSP current is suppressed by a factor of the LSP velocity
$\sim 10^{-3}$.} the maximal possible cross section is thus getting
squeezed by Higgs mass bounds from unsuccessful Higgs searches at LEP
and elsewhere. Constraints from radiative $b \rightarrow s \gamma$ decays
also play a role. As a result, choices of parameters that give a relic
density of the right order of magnitude typically predict \cite{14,52}
LSP detection rates in ${}^{76}$Ge somewhere between $10^{-4}$ and
$10^{-1}$ evts/(kg$\cdot$day), although somewhat higher rates may still
be possible \cite{53}. This means that one will have to improve the
current sensitivity by at least two, and possibly five, orders of magnitude!

The Proceedings of a recent workshop on DM and its direct detection
\cite{54} contains contributions from 11 different groups who are
looking, or are planning to look, for WIMPs. As already mentioned,
the currently best limit comes from NaI scintillator detectors \cite{50}.
Their sensitivity is expected to improve significantly through improved
pulse shape discrimination, doping, and by reducing the operating
temperature \cite{55}. However, among the experiments that are now
running, the CDMS experiment will probably achieve the highest
sensitivity, at least for spin--independent couplings. They just
published \cite{56} first results from a pilot run, where they
operated several Si and Ge detector modules at millikelvin temperatures
at a shallow site in California; the mass of the modules is of order
100 g each. While not yet competitive, this run indicates that their
detectors are working satisfactorily. Once the full experiment is
installed in the Soudan mine, it is expected to improve upon current
sensitivity by at least two orders of magnitude \cite{56}.

This will still not be sufficient to probe much of MSSM parameter space
(although it may see a positive signal). Recently there has been a 
proposal \cite{57} to deploy a large number of enriched ${}^{76}$Ge
detectors, for a total mass of (at least!) one ton, in a big tank of
liquid nitrogen inside the Gran Sasso laboratory. Preliminary studies
indicate that this could indeed cover nearly the entire MSSM parameter
space, for material purities that are worse than what has already been
achieved for the BOREXINO experiment (not for Ge, however). Besides
looking for WIMPs, this experiment could also improve the sensitivity
of searches for neutrinoless double$-\beta$ decays by several orders
of magnitude; in fact, the ``ordinary'' $2 \nu \ \beta \beta$ decay
may well give the biggest background to WIMP searches in this
experiment! Note that the total target mass so far deployed in the CDMS
experiment is less than 1 kg. Clearly 1 ton of ${}^{76}$Ge inside
500 m$^3$ of liquid nitrogen is no longer a ``table top'' experiment, but
it should still be much cheaper to build than a major collider experiment.

One problem of direct WIMP search experiments is that it may be difficult
to convince people that a putative signal is in fact due to WIMPs. Neither
the NaI nor the semiconductor experiments have any directionality, so
at best one can measure a recoil energy spectrum. Since the mass of the
WIMP is not known, no clear prediction for the shape of this spectrum
can be made. Worse, this shape also depends on the assumed velocity
distribution of the WIMPs in the halo. It could, e.g., be changed
significantly if the halo had some net rotation \cite{58}. 

The only fairly robust characteristic WIMP signal that has so far been
suggested is based on the annular modulation of the WIMP flux as seen on
Earth \cite{59}. This signal occurs because the Earth moves around the Sun
with about 30 km/sec. If the DM halo does not rotate in the rest frame
of the (visible) galaxy (the standard assumption), a component of
about 15 km/sec will add to the Sun's velocity around the center of our
galaxy ($v_{\rm sun} \simeq 220$~km/sec) in early June, 
and subtract from it in December. As long as the
rotational velocity of the dark halo does not exceed the orbital velocity
of the Sun around the galactic center, the phase and frequency of this
modulation will not change, although the amplitude might. One thus
expects the counting rate to behave like
\be \label{e20}
N(t) = N_0 + N_1 \cos [ \omega ( t - t_0)],
\ee
where $t_0 = $ June 2 and $\omega = 2 \pi /$yr. Unfortunately $N_1/N_0$
is expected to be quite small ($\sim 5$\% for a non--rotating halo), so
seeing this modulation will not be easy.

Nevertheless the DAMA collaboration announced last year (1997) that they
may have found an effect which is not incompatible with being a signal
\cite{60}. However, as pointed out in ref.\cite{61}, this ``not--bound'' is
almost certainly not a true signal. For one thing, it only occurs in 3
out of 9 crystals; for the other 6 crystals $N_1$ is perfectly consistent
with 0. Also, even the best fit value of $m_\chi$ does not reproduce
the observed energy dependence of the excess very well; however, as
far as I know they have not tried varying the parameters of the galactic
halo model (WIMP velocity distribution and/or bulk halo rotation) in
their fits. Finally, the data were taken either in winter or in June; it
is therefore not surprising that the fit gave a period compatible with
1 yr.

\setcounter{footnote}{0}
\subsection*{3b) Indirect WIMP Detection: Neutrinos}

Of course, WIMPs (if they exist) do not only scatter off nuclei inside
a detector; they can scatter off any nucleus inside the Sun or Earth
(or elsewhere). If they loose a sufficient amount of energy in that
scattering, they will become gravitationally bound to the celestial
body they hit. After (many) additional scatters, they will eventually
spiral into the center of this body. In other words, the WIMP density
should be (greatly) enhanced in the centers of massive celestial bodies.
WIMPs will then begin to annihilate with significant rates inside these
bodies; eventually WIMP capture by and WIMP annihilation in these
bodies will reach equilibrium. In case of the Sun this equilibrium should
have been reached for all reasonable WIMPs, but this may not be true in
case of the Earth if the WIMP mass exceeds 100 GeV or so.

Of course, most particles produced by WIMP annihilation in the Earth or
Sun will get stuck in these bodies. The only particles that have a 
chance to escape are neutrinos. Unfortunately Majorana WIMPs at rest
cannot annihilate into a massless $f \bar{f}$ pair (unless CP is violated
in the WIMP sector, e.g. due to phases in the neutralino mass matrix in
the MSSM \cite{phase}). The neutrinos thus have to come from the decay
of heavier particles. In case of the LSP, the best sources of neutrinos
are usually $\tau^+ \tau^-$ and, for heavier LSPs, $W^+ W^-$ and $ZZ$
final states \cite{14}.

The most easily detected neutrinos are muon (anti--)neutrinos. These
can be searched for with so--called neutrino telescopes. The idea is
that the neutrino converts into a muon through a charged current 
reaction in the material surrounding the detector; the muon itself is
then easily detected, either using conventional muon chambers, or
through the \v{C}erenkov light it emits on its way through (frozen)
water. Due to the large background of cosmic ray induced muons,
one can only hope to see a signal for WIMP annihilation in upwards going
muons. Obviously neutrinos (and hence muons) from WIMP annihilation in
the center of the Earth will always come from below, but in case of
neutrinos from the center of the Sun this requirement reduces the
``duty cycle'' (the useful luminosity) by a factor of two.

Note that in equilibrium the annihilation rate is half the capture
rate, which in turn is given by the WIMP--nucleus scattering cross
section. The expected rates for direct and indirect WIMP detection
are therefore correlated \cite{62}. However, the dependence on the
WIMP mass is different for the two signals. In case of direct
detection, if the WIMP mass is less than or of order of the mass of
the target nucleus, increasing $m_\chi$ increases the average recoil
energy, eq.(\ref{recoil}), which makes these events easier to
detect.\footnote{However, increasing the energy region where a signal
can exist might also increase the total background.} The increase of
the available phase space also increases the total scattering cross
section. However, for large WIMP masses the direct detection rate
drops like $1/m_\chi$, due to the decrease in flux, eq.(\ref{e5}).
On the other hand, both the $\nu_\mu \rightarrow \mu$ conversion
probability and the length of the muon track (which determines the
effective detector volume) increase more or less linearly with the
average neutrino energy, which in turn is proportional to the WIMP
mass. One would therefore naively expect the indirect detection rate
to increase linearly with $m_\chi$, even after the reduction of the
flux (\ref{e5}) has been taken into account.\footnote{Note also that
the angular correlation between the $\mu$ and $\nu_\mu$ directions
becomes better at higher energies. Heavier WIMPs would therefore
give sources with smaller opening angles, and hence less background.
The main physics background comes from atmospheric neutrinos.}
However, heavier WIMPs are less likely to become gravitationally
bound after they scatter \cite{14}. Moreover, at least in case
of MSSM neutralinos, heavier LSPs are usually less mixed; note that
the $\lsp \lsp$Higgs couplings are proportional to the product of
higgsino and gaugino components of \lsp. Heavier LSPs hence tend to
have lower scattering cross sections. The combination of these competing
effects means that in the MSSM there is no strong correlation between
the expected indirect detection rate from LSP annihilation in the
Earth or Sun and the LSP mass \cite{63}.

The best bounds on the flux of upward--going muon neutrinos currently
come from the Baksan collaboration \cite{64}. Their detector consists
of several layers of muon chambers covering $\sim 250$~m$^2$; it has
been taking data for about 20 years. This bound gives the best
constraint on heavy Majorana neutrinos as DM candidates \cite{kami},
but it only begins to scratch the MSSM parameter space \cite{63}. One
would need to improve sensitivity by 4 or 5 orders of magnitude to
probe most of parameter space. The required large increase in detector
area is probably only affordable if one instruments a large volume of
(frozen) water with photo--multiplier tubes (PMTs), and searches for
muon tracks through their \v{C}erenkov light. The so far most
promising experiment along these lines is AMANDA, which has been
operating successfully at the South Pole for a couple of years
\cite{65}. They have now demonstrated their ability to distinguish
upward going (neutrino induced) events from cosmic ray backgrounds,
but currently their threshold energy is still quite high, $E_\mu^{\rm
thr} \simeq 50$~GeV; this limits their usefulness for WIMP searches
since typically $\langle E_\mu \rangle \leq m_\chi / 4$. Currently the
effective detector area of AMANDA is about $10^4$~m$^2$; they
eventually plan to instrument an entire km$^3$ of ice. Since deep
polar ice has proven to be exceptionally transparent, about 5,000 PMTs
would be sufficient for this ambitious experiment \cite{65}. At
present it is not clear what the ultimate energy threshold of this
``ICE CUBE'' detector would be. There are also several proposals to
deploy \v{C}erenkov detectors in deep lake or sea water \cite{66}.

The energy resolution of most existing or planned neutrino telescopes is
quite poor. The only exception I know of is the HANUL experiment in
Korea \cite{hanul}, which plans to use large permanent magnets as
muon spectrometers. Excellent up--down rejection (required to suppress
cosmic ray muon backgrounds) is to be achieved by precise timing, as well
as by PMT--instrumented water tanks above and below the magnets (which
determine the direction of the \v{C}erenkov cones of the muons, and hence
their flight direction). If everything works as planned, this might allow
one to do neutrino astronomy above ground. However, the currently foreseen
number of modules is probably too small to detect a signal for LSP
annihilation.

\setcounter{footnote}{0}
\subsection*{3c) Indirect WIMP Detection: Annihilation in the Halo}

Even though the WIMP density in the galactic halo is expected to be
far lower than that in the center of celestial bodies, halo WIMPs should
still annihilate occasionally. In this case all annihilation products
are in principle visible, although it is not a priori clear whether
they will be detectable on top of the expected backgrounds. Three
channels appear to have some potential for WIMP detection \cite{14}:
positrons, antiprotons, and $\gamma$ lines.

Since positrons are very light, they can again not be produced in
Majorana WIMP annihilation at rest, but are expected to originate from
the decay products of heavier particles. In particular, annihilation
into $W$ and $Z$ pairs might give rise to a prominent feature in the
$e^+$ energy spectrum at $E_{e^+} \sim m_\chi/2$ \cite{14}. In
contrast, the background is expected to be a smoothly falling
function of the positron energy.

Antiprotons can originate from hadronic decays of WIMP annihilation
products. Since these antiprotons will be produced at the end of a
fairly long decay and hadronization chain, their momentum is expected to
be only a small fraction of the WIMP mass. In contrast, most background
antiprotons come from the interaction of energetic cosmic rays with
more or less stationary targets, and are therefore expected to be
quite energetic due to a Lorentz boost. The best signal for WIMP annihilation
in this channel is therefore expected to show up at low energies, well
below 1 GeV \cite{14}.

Both $e^+$ and $\bar p$ have to be detected near the top of or above the
atmosphere. A significant improvement of the current $\bar p$ flux
measurements (from balloon experiments) is expected to be achieved by
the Alpha Magnetic Spectrometer \cite{ams}, which is scheduled to be
flown on a Space Shuttle this year, and should eventually be installed
on the international space station (assuming it ever gets built).

Predictions for both these signals suffer from uncertainties in the
modeling of charged particle propagation through the galaxy. Charged
particles in the expected energy range will get trapped by the magnetic
field of the galaxy; they may thus have spent a lot of time orbiting the
galaxy before reaching Earth. In case of the $\bar p$ signal the energy
is so low that effects of the solar $B-$field, including its time
dependence, also have to be modeled carefully. 

These complications do not arise for the $\gamma$ line signal for WIMP
annihilation in the halo, since the produced photons simply travel in a
straight line (more exactly, a geodesic). On the other hand, WIMPs by
definition have no tree--level couplings to photons. WIMP annihilation
into $\gamma \gamma$ or $\gamma Z$ can therefore only occur through
loops. The first complete calculations of the corresponding matrix elements
in the MSSM have been completed only last year \cite{67}. For
gaugino--like LSPs the resulting cross section is small, much less than
1\% of the total annihilation cross section. On the other hand, these
loop--induced cross sections become quite important for heavy
higgsino--like LSPs. Indeed, refs.\cite{67} find constant 
($m_\chi-$independent) cross sections in this case, in conflict with partial
wave unitarity. Since the calculation for the $\gamma \gamma$ final state
has been performed by two independent groups, their result is probably
correct. However, multi--loop effects should unitarize the cross
section, i.e. restore the required $1/m^2_\chi$ behavior. The results
of refs.\cite{67} violate unitarity only for $m_\chi \geq 100$~TeV,
which is of no practical interest; however, already for $m_\chi \sim 2$~TeV
the loop--induced $\lsp \lsp \rightarrow \gamma \gamma$ cross section
comes out {\em larger} than the cross section for $\chi_1^+ \chi_1^-
\rightarrow \gamma \gamma$, which occurs at tree--level. I find it hard
to believe that this result will survive in a more complete
calculation.\footnote{Last year there has been a claim \cite{68} that
data from a balloon emulsion experiment and from air shower \v{C}erenkov
telescopes show evidence for a line at 3.5 TeV; however, this claim 
seems to have been based on an incorrect interpretation of the data
\cite{69}. Besides, this ``signal'' would have been far stronger than
what one expects in any known model.}

The strength of the signal for WIMP annihilation into $\gamma \gamma$ and
$\gamma Z$ final states is proportional to the square of the WIMP 
density, integrated along the line of sight in the direction one is
looking. If WIMPs exist, they should be most abundant near the center
of our galaxy. Unfortunately the value of the DM density near the galactic
center is much less constrained by observation than the local DM
density is. Predictions for the strength of this signal using different
halo models therefore differ by several orders of magnitude \cite{70}.
If the halo density is quite singular near the center, prospects for
detecting heavy higgsino--like LSPs in this channel may be quite
good \cite{70}; recall, however, that higgsinos need to be
uncomfortably (from the particle physics point of view) heavy to make
good DM candidates. If the LSP is gaugino--like, and/or the halo
density varies smoothly near the center of the galaxy, the $\gamma$
line signal does not appear to be very promising.

We thus see that predictions for all signals for WIMP annihilation in
the galactic halo suffer from large astrophysical (or
galacto--physical) uncertainties. It will therefore be difficult to
translate null results into bounds on parameter space, or positive
observations into measurements of cross sections, until we know much
more about the dynamics of our galaxy. For the time being the signals
discussed in this subsection should therefore be considered to be
potential discovery channels, rather than as ways to possibly rule
out certain DM candidates.

\section*{4) Summary and Conclusions}

Most of the mass of the Universe is dark. Very likely most of this
Dark Matter is non--baryonic, although baryonic Dark Matter should
also exist, and may have been found in the form of MACHOs. 
Neutrinos, by themselves, do not appear to make good DM candidates;
the DM puzzle therefore strongly hints towards new physics. 

Unfortunately knowing the approximate DM density does not allow us to
say much about the objects that form it. Even if we restrict ourselves
to truly elementary particles, their mass could be anywhere between
$10^{-5}$~eV (axions; see Sec.~2b) and $\geq 10^{13}$~GeV (Sec.~2f).
Their interactions with normal (baryonic) matter could be anywhere
between essentially non--existent (gravitinos; Sec.~2e) and
extremely violent ($B-$balls; Sec.~2g).

Out of this zoo of exotic hypothetical DM candidates, the supersymmetric
LSP (Sec.~2c) remains my personal favorite. The agreement (to
logarithmic accuracy) between expected and ``observed'' relic density,
especially for gaugino--like LSPs, seems too close to be a mere
coincident. Also, supersymmetry is well motivated quite independent
of the DM problem. (However, axions also have a, to my mind weaker, particle
physics motivation; and some SUSY models allow for other DM candidates.)
The detection of relic neutralinos, while probably far from easy, is at
least not hopeless (Sec.~3). I would be surprised if the necessary
sensitivity will be achieved before the turn--on of the LHC experiments,
which are almost guaranteed to detect superparticles if they exist
``at the weak scale''. However, even if SUSY is first discovered at
colliders, detecting relic neutralinos (or other DM particles) remains
of the greatest importance. For one thing, collider experiments will
never be able to prove that the LSP is sufficiently stable to form
DM (although they might eliminate it from the list of candidates if
it is very short--lived); the best bound on the LSP lifetime from
these experiments is expected to be some 24 orders of magnitude
less than the age of the Universe. Conversely, once the mass and
interaction strength of the LSP (or any other DM candidate) have been
determined by collider experiments, and its longevity has been
established by any one positive DM signal, various (bounds on) 
``indirect'' signals for WIMP annihilation will allow us to constrain
galactic models; a major source of uncertainty in attempts to understand
the formation of large scale structures in the Universe will then also
have been removed. Accelerator--based experiments and Dark Matter searches
should therefore be considered to complement, rather than compete with,
each other.

\subsection*{Acknowledgements}
I thank the ICTP Trieste for their financial support, which allowed me
to attend this exceptionally well organized meeting.

\end{document}